\documentclass[graybox]{svmult}

\usepackage{mathptmx}       
\usepackage{helvet}         
\usepackage{courier}        
\usepackage{type1cm}        
\usepackage{float}                            
\usepackage{makeidx}         
\usepackage{graphicx}        
\usepackage{multicol}        
\usepackage[bottom]{footmisc}
\usepackage{amsmath} 
\usepackage{color}
\usepackage[babel,german=guillemets]{csquotes}
\definecolor{keywordcolor}{rgb}{0,0,1}

\makeindex             

\begin{document}

\title*{Single Semiconductor Quantum Dots in Microcavities:
Bright sources of indistinguishable Photons
}
\author{C. Schneider, P. Gold, C.-Y. Lu, S. H\"ofling, J.-W. Pan, M. Kamp}
\institute{C. Schneider \at University of W\"urzburg, Am Hubland, W\"urzburg, Germany \email{Christian.schneider@physik.uni-wuerzburg.de}}
%
%
\maketitle


\abstract{In this chapter we will discuss the technology and experimental techniques to realize quantum dot (QD) single photon sources combining high outcoupling efficiencies and highest degrees of non-postselected photon indistinguishability. The system, which is based on ultra low density InAs QDs embedded in a quasi planar single sided microcavity with natural photonic traps is an ideal testbed to study quantum light emission from single QDs. We will discuss the influence of the excitation conditions on the purity of the single photon emission, and in particular on the degree of indistinguishability of the emitted photons. While high purity triggered emission of single photons is observed under all tested excitation conditions, single photon interference effects can be almost vanish in experiments relying on non-resonant pumping into the quantum dot wetting layer. In contrast, we can observe nearly perfect indistinguishability of single photons in resonance fluorescence excitation conditions, which underlines the superiority of this excitation scheme to create photon wave packets close to the Fourier limit. As a first step towards the realization of solid state quantum networks based on quantum dot single photon sources we test the overlap of photons emitted from remote QDs yielding non-postselected interference visibilities on the order of ($\approx 40 \%$) for quasi resonant excitation.
}  
\clearpage
\indent

\section{Introduction}

The demonstration of single photon emission from a semiconductor quantum dot (QD) \cite{Michler-Science00} triggered 15 years of prospering research devoted towards the application of semiconductor quantum light emitters. In particular the most commonly studied InGaAs/GaAs based QDs can emit highly non-classical light, both under optical and electrical excitation \cite{Yuan-Science02, Michler-Science00, Heindel-APL10, Michler-1}. However, for many applications beyond simple quantum cryptography schemes, it is vitally important that the emitted single photons are comprising highest degrees of indistinguishability, which means that they have to be identical in all spectral characteristics: Their color, polarization, and furthermore the extension of the wave packet (i.e. their coherence) should be Fourier limited \cite{Michler-2, Santori-Nature02}. Such photons are at the heart of applications in quantum communication \cite{Pan-RMP12}, quantum networks \cite{Kok-RMP07} and linear optical quantum computing \cite{Obrien-Science07}. Most quantum teleportation schemes strongly rely on photon indistinguishability, and in particular the route towards quantum repeater networks for future long distance quantum communications highly relies on this property \cite{Briegel-PRL98, Hofmann-Science12, Cirac-PRL97}. First important experiments relying on quantum interference and teleportation of photons emitted from single QDs have recently been carried out \cite{Nilsson-NatPhot13, Gao-NatCom13}. In contrast to isolated quantum emitters such as atoms in dilute vapors, QDs are embedded in a solid state environment which imposes limitations on 
\begin{itemize}
	\item the brightness of the source since the photons have to be extracted out of a high refractive index material
	\item  the interference properties of photons emitted from these sources, as coherence and color of the emission can be affected by coupling to the environment of the emitter.
\end{itemize}
 While the source efficiency can be boosted to very high values by embedding quantum emitters in photonic micro- and nanostructures \cite{Heindel-APL10, Gazzano-NatCom13, Claudon-NatPhot10, Reimer-NatCom12}, increasing the degree of indistinguishability at least partly requires to decouple the emitter from its environment. In particular frequency shifts induced by charges in the QD's vicinity, or effects of phonon induced emitter dephasing strongly and detrimentally affect the interference properties of single QDs. 
This chapter is structured as follows: In section~\ref{section:theory} we will briefly address the fundamental mechanisms of two photon interference and its experimental implementation. In section~\ref{section:SPS} we will describe the experimental realization of a bright single photon source as the basis of the following studies. Finally, in sections~\ref{section:experimentI} and \ref{section:experimentII} we will describe two photon interference experiments carried out on single QDs. We will assess in detail the influence of the excitation scheme on the interference properties of the quantum light emitted from the QD.  While carrier refilling effects are identified to strongly detrimentally influence the photon interference in non-resonant excitation schemes, resonance fluorescence conditions can lead to almost perfect interference visibilities of photons. We can furthermore demonstrate significant interference of single photons emitted from separate sources and compare our experimental findings with an analytical model.

\section{A pedestrian's guide to two photon interference}
\label{section:theory}

\subsection{Quantum dot single photon source}

In a very simplified picture, single QDs can be considered as two level systems embedded in a solid state environment: Electrons and holes can be captured and localized in QDs if the band configuration provides an potential well in the conduction and valence band, and the small size of the QDs leads to strong Coulomb and exchange interactions. As a result, the energetic ground state of a single dot can only be occupied by a single electron-hole pair. This excitonic carrier complex can decay spontaneously and its energy is transferred to a single photon. Since the maximum occupation number of the excited state cannot exceed unity, not more than one photon of the corresponding energy can be emitted at a time interval on the order of the exciton lifetime. The distribution function of the photon stream is hence sub-poissonian (or antibunched), which is usually characterized by the second order correlation function which we write here in terms of the emitted photon intensities and the delay time $\tau$ between two detection events: $g^{(2)}(\tau)=\frac{<I(t)I(t+\tau)>}{<I(t)>^2}$. For an ideal single photon source, the value of this function approaches 0 at $\tau=0$. The antibunched nature of the emission from a single QD makes these quantum emitters highly interesting for quantum cryptography schemes relying on encoding the information of a quantum key into the polarization of a single light particle (such as the famous BB84 protocol \cite{Bennett-84}). First successful experimental demonstrations of quantum key distribution with QD single photon sources, both under optical \cite{Waks-Nature02} and electrical excitation \cite{Heindel-NJP12} have been realized. However, for more advanced schemes such as the remote entanglement of stationary quantum bits (Qubits), the interference properties of these photons play a dominant role, which we will detail in the following subchapter.

\subsection{Photon interference with quantum light}

\begin{figure}[h]
\centering
\includegraphics[width=12cm]{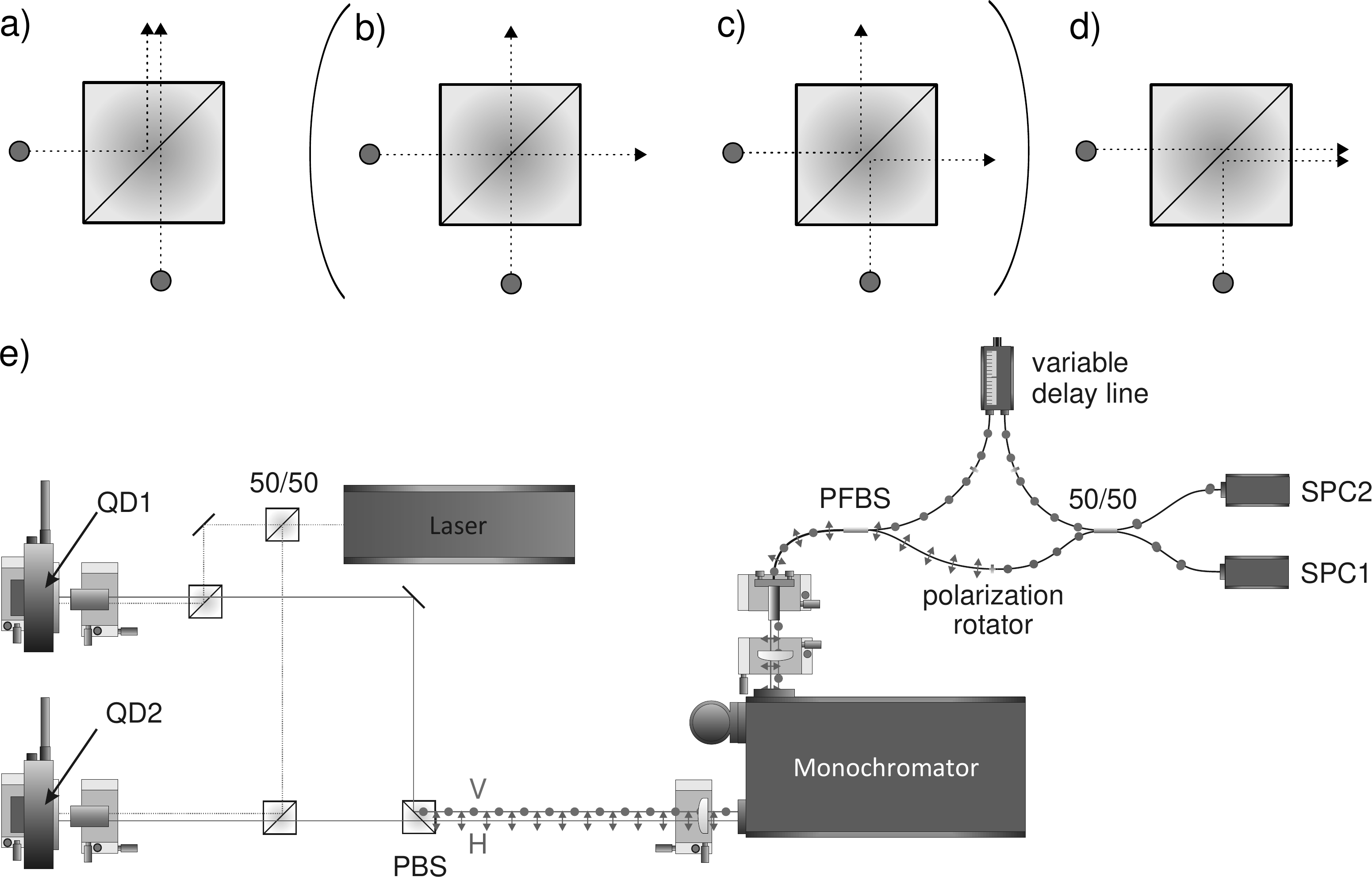}
\caption{\label{Abb:HOM} a-d) Possible configurations of two photons entering a beam splitter from different sides. Quantum interference determines the output paths of the photons. If the two input photons were indistinguishable, both photons can only leave the beam splitter in pairs through a common arm (a and d), rather than separate arms (b and c). e) Experimental implementation of the two photon interference experiment: The photons are emitted either from the same source (QD1), or from two separate sources (QD1 and QD2) mounted in different cryostats. In the latter case, the photons are brought together on a polarizing beam splitter (PBS). After spectral filtering, the light is fed into an asymmetric Mach-Zehnder interferometer via a polarizing fiber beam splitter (PFBS), and the two photon interference effect occurs on the last 50/50 beam splitter. Single photon counters (SPCs) are connected to each exit port of the beam splitter to record quantum correlations. 
}
\end{figure}

The interference properties of single photons and their indistinguishability are very closely related properties. Indistinguishable photons share all relevant properties, such as color, polarization, and extension of the wave-packet in time. 
Directly probing the indistinguishability of single photons is usually carried out in the interference experiment pioneered by Hong, Ou and Mandel \cite{HOM-PRL87}. For further details see also the chapter (by Kuhn, Zhao). It is manifested as a quantum interference effect when two photons arrive from different sides on a beam splitter: If these photons are indistinguishable and if they overlap in space and time (spatio-temporal overlap) on the beam splitter, quantum interference will force them always to exit through a common output port, which creates a path entangled  (N=2) NOON-state. This is schematically sketched in Fig.~\ref{Abb:HOM}a and d). The fact that the photons leave the beam splitter in bunches with a suppression of the scenarios sketched in ~\ref{Abb:HOM}b and c) as a result of destructive quantum interference reflects their bosonic nature, and cannot be explained by classical electrodynamics.

The Hong-Ou-Mandel effect can be experimentally probed by utilizing a configuration of single photon detectors similarly to a Hanbury Brown and Twiss setup which is routinely used to probe the quantum nature of the emitted light \cite{Michler-1}. The setup is schematically sketched in Fig.~\ref{Abb:HOM}e). The photons can be emitted from a single source, or from separate, distant sources mounted in separate cryostats. In the latter case, the photon beams are brought together on a polarizing beam splitter and fed into an unbalanced Mach-Zehnder fiber interferometer after spectral filtering. One arm of the interferometer has a tunable length to adjust the arrival time of the photons on the second beam splitter, where the two photon interference is probed. Single photon detectors are connected to each output port of the beam splitter. The second order photon correlation function is recorded by measuring the time delays between the measurement events of the individual detectors. In case that the photons always take a common exit on the beam splitter, no coincidence detections between both APDs occur, and the second order correlation function approaches 0 at $\tau=0$. Here, we have to remember that the single photons emitted from our QD source are photon wave packets, which interfere on the beam splitter. Ideally, these wave packets are Fourier-transform limited, with a Lorentzian spectral broadening $\Delta \omega$ being solely determined by the emitter decay time $\tau_r$, and the temporal extension of the wave packet is given by $\tau_c=2 \times \tau_r$. If additional dephasing channels with a characteristic time $\tau_{deph}$, such as coupling to phonons start to play a role, the coherence time $\tau_c$ is reduced to $\frac{1}{\tau_c}=\frac{1}{2\tau_r}+\frac{1}{\tau_{deph}}$. For Fourier-transform limited wave packets, the second order correlation function should reach a value of 0 at $\tau=0$, which converts into a two photon indistinguishability of 100$\%$. To understand the shape of the second order correlation function around $\tau=0$ in the presence of dephasing, one has to calculate the overlap integral of two photons incident on the beam splitter. The correlation function of the $\tau=0$-peak for two photons with identical frequency and a time delay $\delta\tau$ between them is given by \cite{Bylander-EPJD03}

\begin{equation}\label{eq:g2hom}
g^{(2)}(\tau, \delta\tau)= \frac{1}{4} e^{-\frac{\left|\tau-\delta\tau\right|}{\tau_r}}+\frac{1}{4} e^{-\frac{\left|\tau+\delta\tau\right|}{\tau_r}}-\frac{1}{2}e^{-\left(\frac{2}{\tau_c}-\frac{1}{\tau_r}\right)\left|\tau\right|-\frac{\left|\tau-\delta\tau\right|}{2\tau_r}-\frac{\left|\tau+\delta\tau\right|}{2\tau_r}}.
\end{equation}

In equation (\ref{eq:g2hom}) a homogeneous broadening is assumed by an exponential decay of the coherent amplitude due to dephasing and the existence of (gaussian) spectral jitter is ignored. 

\begin{figure}
	\centering
\includegraphics[width=11cm]{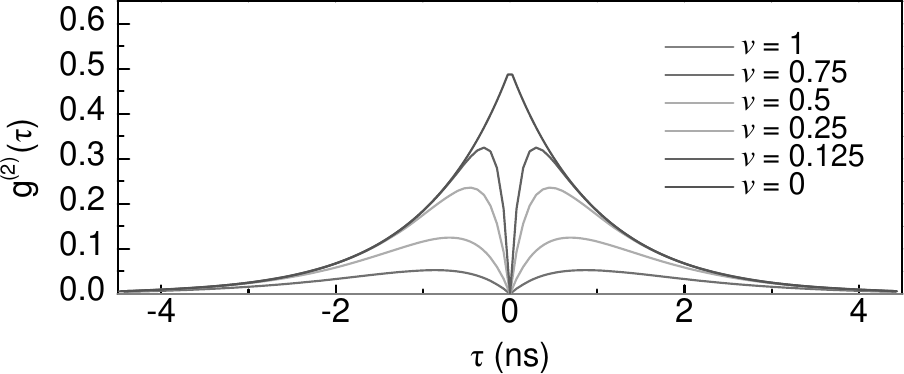}
	\caption{Calculated correlation function $g^{(2)}(\tau)$ of the $\tau=0$-Peak for the interference of indistinguishable photons with a radiative decay time of $\tau_r=1$ ns for different visibilities $v=\tau_c/2\tau_r$.}
	\label{fig:Fig2}
\end{figure}

The correlation function  $g^{(2)}(\tau)$ of the $\tau=0$-peak for the interference of photons with the same energy and polarization from a pulsed source is shown in Fig.~\ref{fig:Fig2} for different ratios of $v=\tau_c/2\tau_r$. We calculated $g^{(2)}(\tau)$ for a time delay of $\delta\tau=0$ and a constant radiative decay time of $\tau_r=1$ ns. The variation of $v$ is achieved by varying the coherence time $\tau_c$ between 2 ns and 0 ns. In the case of Fourier-tranform limited photons the visibility is 100\% and the $\tau=0$-peak disappears completely. For a homogeneous broadening of the emission, represented by a coherence time $\tau_c<2\tau_r$, the wavepackets do not overlapp perfectly which results in a non-zero contribution to the correlation funtion around $\tau=0$. In the limit of very short coherence times $\tau_c\rightarrow0$, the photons leave the beam splitter independently and randomly resulting from the reduced coalescence probability. The outcome of this is a peak in the correlation histogram with a $g^{(2)}(0)$ value of 0.5, which equals the $g^{(2)}(0)$-value for a two photon source $1-1/n$ with $n=2$.

\section{A bright quasi-planar single photon source}
\label{section:SPS}

\begin{figure}[h]
\centering
\includegraphics[width=12cm]{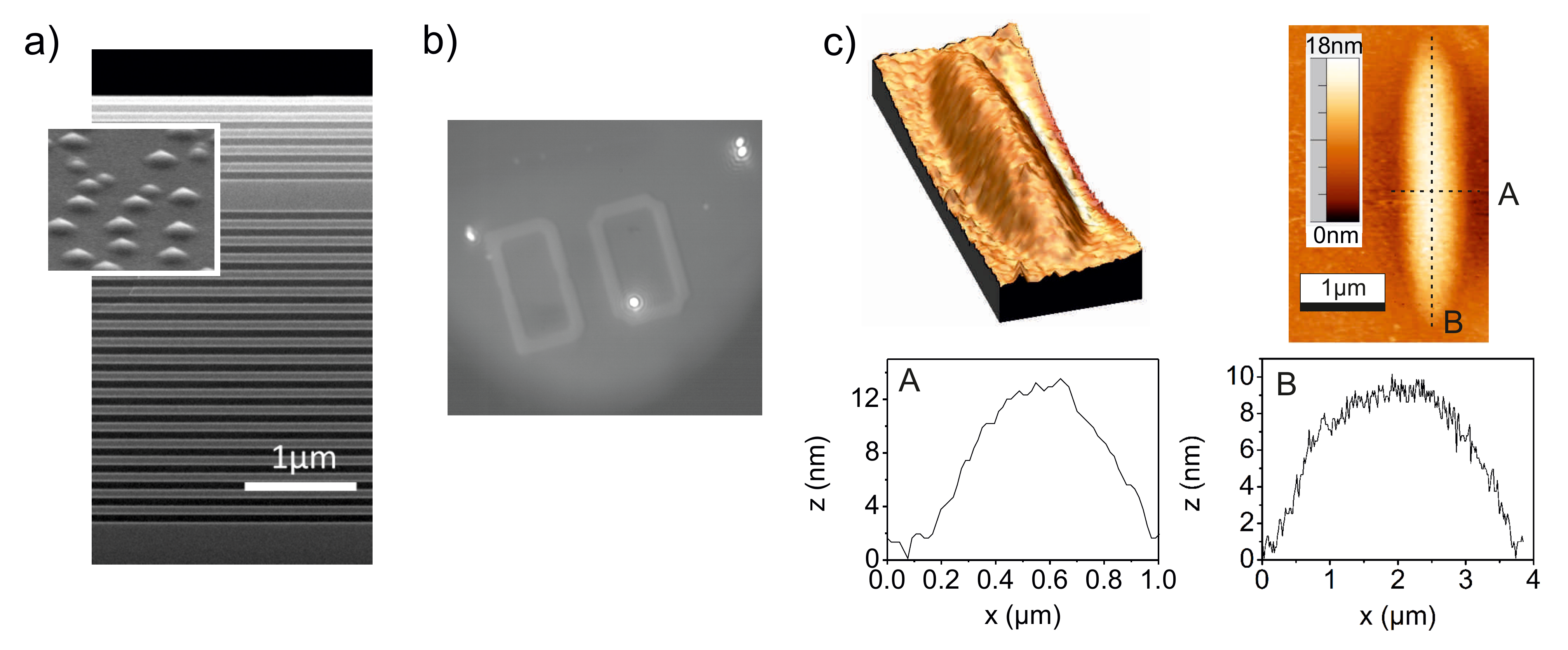}
\caption{\label{Abb:SPS1} a) Sample structure and quantum dots: A single layer of In(Ga)As QDs is integrated in a single sided low-Q cavity realized by AlAs/GaAs distributed Bragg reflectors sandwiching a GaAs cavity layer. b) The very low QD density facilitates the identification of QD emission spots straight forwardly by white light imaging. c) Nanohills on the sample surface are formed during the growth process acting as natural lenses and significantly improve light outcoupling of the structure. The figure is reproduced from Maier et al. \cite{Maier-OE14}. 
}
\end{figure}

The single photon source which is at the heart of this study is based on a low density layer of single In(Ga)As QDs integrated in an asymmetric AlAs/GaAs optical microcavity. The lower distributed Bragg reflector (DBR) consists of 18 AlAs/GaAs mirrorpairs, providing a reflectivity near unity in the spectral range of the QD emission, which directs the light towards the top surface of the structure. A single layer of ultra low density InAs QDs is vertically centered in a GaAs $\lambda$-thick cavity layer, which is covered by 5 AlAs/GaAs distributed Bragg reflector (DBR) segments. The microcavity has a low quality factor (Q-factor) of $\approx$ 200. A combination of very low growth rates ($<$ 0.01 nm/s), and the partial capping and annealing growth technique \cite{Garcia-APL97} allows us to realize sufficiently low QD densities to spectroscopically isolate single QDs in the wavelength range between 900 nm and 940 nm. It is interesting to note that under such growth conditions, which allow for very long migration lengths of the supplied material, the QDs tend to nucleate at crystal steps, defects or nanoholes \cite{Schmidt-Buch, Michler-1}. This peculiar nucleation behavior can be directly exploited in QD positioning schemes \cite{Schmidt-Buch, Schneider-Nanotechnology09}, where nanodefects are intentionally generated on a surface via lithography and etching. In our case, the natural formation of oval crystal defects, which was most likely induced by Gallium droplets during the growth of the bottom DBR similar to observations in \cite{Zajac-PRB12}, serves as such nucleation sites for the QDs in the cavity layer. These defects propagate through the top DBR and are well detectable as nanohills on the the surface (Fig.~\ref{Abb:SPS1}c) via atomic force microscopy, with a height on the order of 10 nm. 
Fig.~\ref{Abb:SPS1}b) depicts a CCD image of the sample surface under illumination with white light at a sample temperature of 4K. We used a long pass filter (750 nm) to monitor the emission from the QDs in the infrared range. The image is characterized by bright spots which we attribute to the emission of clusters of QDs, whereas no detectably signals occur between these sites.

Comparing the position of bright photoluminescence spots recorded via spatially resolved sample imaging with the nanohill position reveals a coalescence between the position of these hills and the location of QDs in the cavity. Furthermore, the oval shape of the hills provides a gentle optical lateral confinement \cite{Zajac-PRB12} which serves to guide the emitted light out of the semiconductor structure and enhance the photon outcoupling efficiency of the device. For a perfectly two dimensional microcavity structure of our geometry, this efficiency can hardly exceed theoretical values of $\approx 30\%$ when the light is collected in the normal direction with a 0.7 NA microscope objective \cite{Royo-JAP01}. In contrast, due to the waveguiding effect provided by the nanodefect, this efficiency can be theoretically increased to $\approx 50\%$, as described in \cite{Maier-OE14}. It is worth noting, that more carefully designed shapes of buried Gaussian nanohills are predicted to facilitate strong mode confinement to the sub-micrometer range without strongly reducing cavity Q-factor via lateral scattering losses \cite{Ding-PRB13}, which makes them very appealing for cavity quantum electrodynamic experiments. In our structure, we could experimentally determine single photon outcoupling efficiencies up to 42 $\%$ in a calibrated photoluminescence setup, being in good agreement with the numerical estimations based on realistic sample parameters, and exceeds the theoretical maximum of a perfectly planar two dimensional microcavity ($\approx 30\%$) \cite{Maier-OE14}. Most other strategies to achieve bright single photon emission from QDs embedded in a semiconductor matrix are based on the integration in nano- and microphotonic structures, such as pillar microcavities, photonic crystal membranes, nano-waveguides and antennas \cite{Barnes-EPJD02}. In all those approaches, the lithographic definition of the photonic structure creates open surfaces in the close vicinity to the quantum emitter, which can lead to significant dephasing and spectral wondering of the QD emission line. This is partly reflected in emitter line broadening \cite{Claudon-NatPhot10, Reimer-NatCom12}, limitations of the QD single photon interference properties \cite{Gazzano-NatCom13} and fast spin dephasing \cite{Press-Naturephotonics10}.

\section{Emission of single and indistinguishable photons from single quantum dots}
\label{section:experimentI}
\subsection{Single photon emission from single QDs}

\begin{figure}[h]
\centering
\includegraphics[width=12cm]{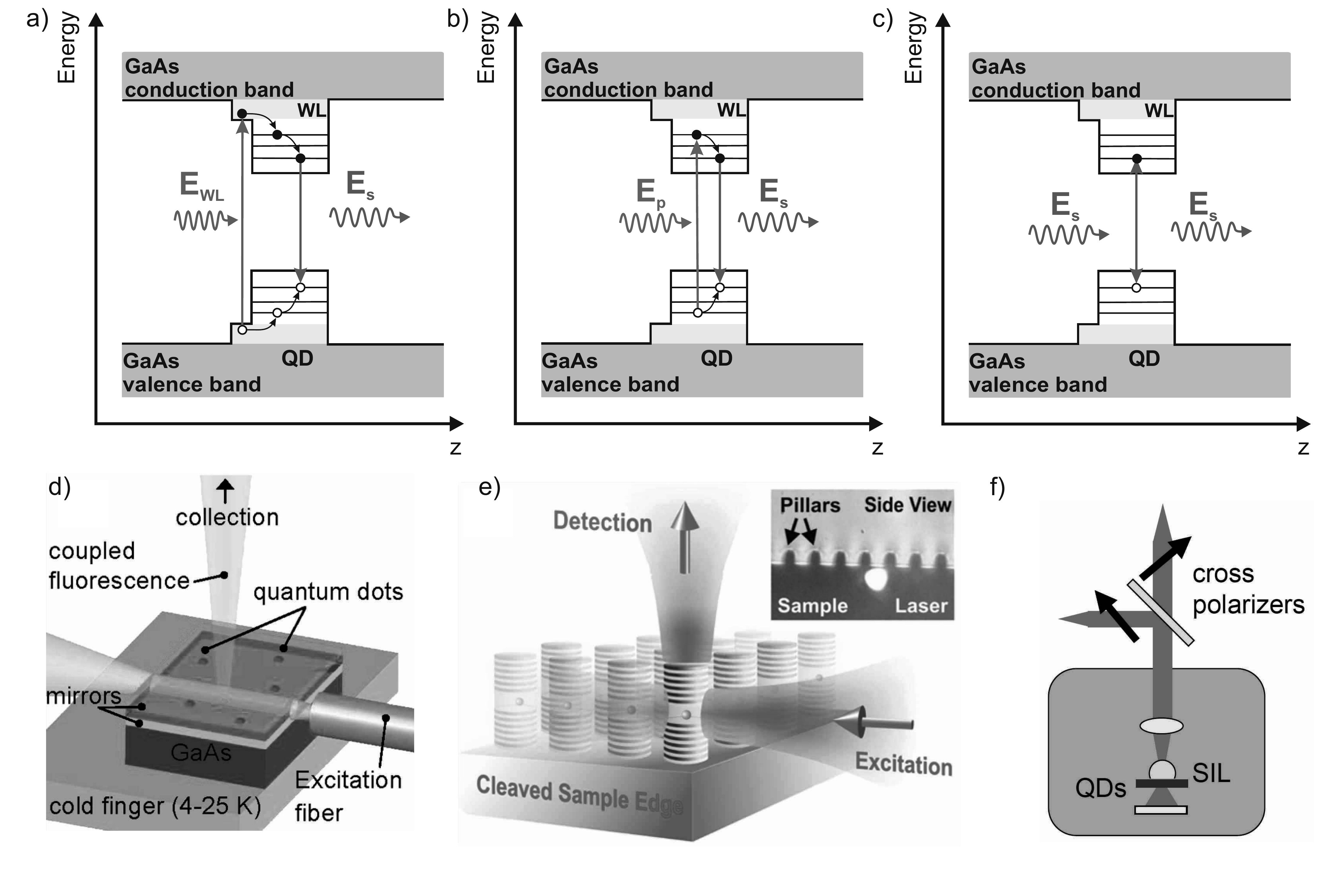}
\caption{\label{Abb:pumping-sketch} Excitation configurations of a single QD: a) Non-resonant wetting layer excitation creates carriers in the surrounding of the QD, which relax into the ground state by scattering. b) Quasi-resonant excitation leads to a direct excitation of the QD without generating a carrier reservoir. c) Resonance fluorescence selectively excites the QD ground state. For suppression of back scattered laser light in resonance fluorescence, the pathways for excitation and photon collection can be separated in the experimental implementation by an orthogonal arrangement for planar waveguides (d) and micropillars (e). f) Sketch for the cross-polarization configuration facilitating resonance fluorescence measurements under normal incidence. d) and (e) are reproduced from Muller et al. \cite{Muller-PRL07} and Ates et al. \cite{Ates-PRL09}. }
\end{figure}

There have been three different optical excitation method used for single-photon generation from QDs, as sketched in Fig. ~\ref{Abb:pumping-sketch}. The most conventional method is non-resonant excitation by pump laser with an energy above the band gap (in the barrier or the wetting layer surrounding the QDs, sketched in Fig.~\ref{Abb:pumping-sketch} a). These carriers can then be captured by the QD and relax to the ground state via phonon scattering, from where they can decay radiatively. In order to facilitate local generation of excitons in the QD and to reduce possibly detrimental effects from the surrounding, carriers can as well be generated quasi-resonantly in the excited states of the QD (Fig.~\ref{Abb:pumping-sketch} b). These states are typically located 20-50 meV on the high-energy side of the exciton ground state in the QD, facilitating spectral filtering of the excitation laser from the collected signal in most cased. The relaxation from the p-shell to the ground state of the QD typically occurs on the ten picoseconds scale \cite{Flagg-PRL12, Santori-Nature02}, which leads to a strong reduction of time jitter in the emission. The third pump-configuration is strictly resonant excitation (resonance fluorescence) a more controlled method widely used in standard atomic physics experiments (Fig.~\ref{Abb:pumping-sketch}c). Here, the excitation laser  tuned on resonance with the QD transition coherently excites the QD. This excitation condition is by far the hardest to implement, since it requires a careful distinction between QD signal and laser stray light. Spectral filtering of the pump-laser is no longer feasible in strictly resonant excitation conditions, hence other methods have to be applied to isolate the QD emission signal: spatial filtering is one option, when the pump-laser is exciting the QD in the perpendicular direction to the collection beam-path \ref{Abb:pumping-sketch}d). This technique was first employed in Bragg waveguides and led to the first successful demonstration of resonance fluorescence from a single QD \cite{Muller-PRL07, Flagg-NatPhys2009}. Strictly resonant excitation perpendicular to the photon collection direction has also been carried out on single micropillar cavities with sufficiently strong suppression of scattered laser light facilitating two photon interference studies \cite{Ates-PRL09}. Another technique is polarization filtering: Here, the excitation laser is linearly polarized, and in case the reflected laser beam preserves this polarization, the QD emission can be detected in the perpendicular polarization basis \cite{Vamivakas-NatPhys09}. In our confocale microscope setup, the polarization extinction of the pump laser can reach values in excess of $10^7$ which allows us to carry out correlation measurements in the resonance fluorescence configuration \cite{He-NatNano13}.  \\

\begin{figure}[H]

\includegraphics[width=12cm]{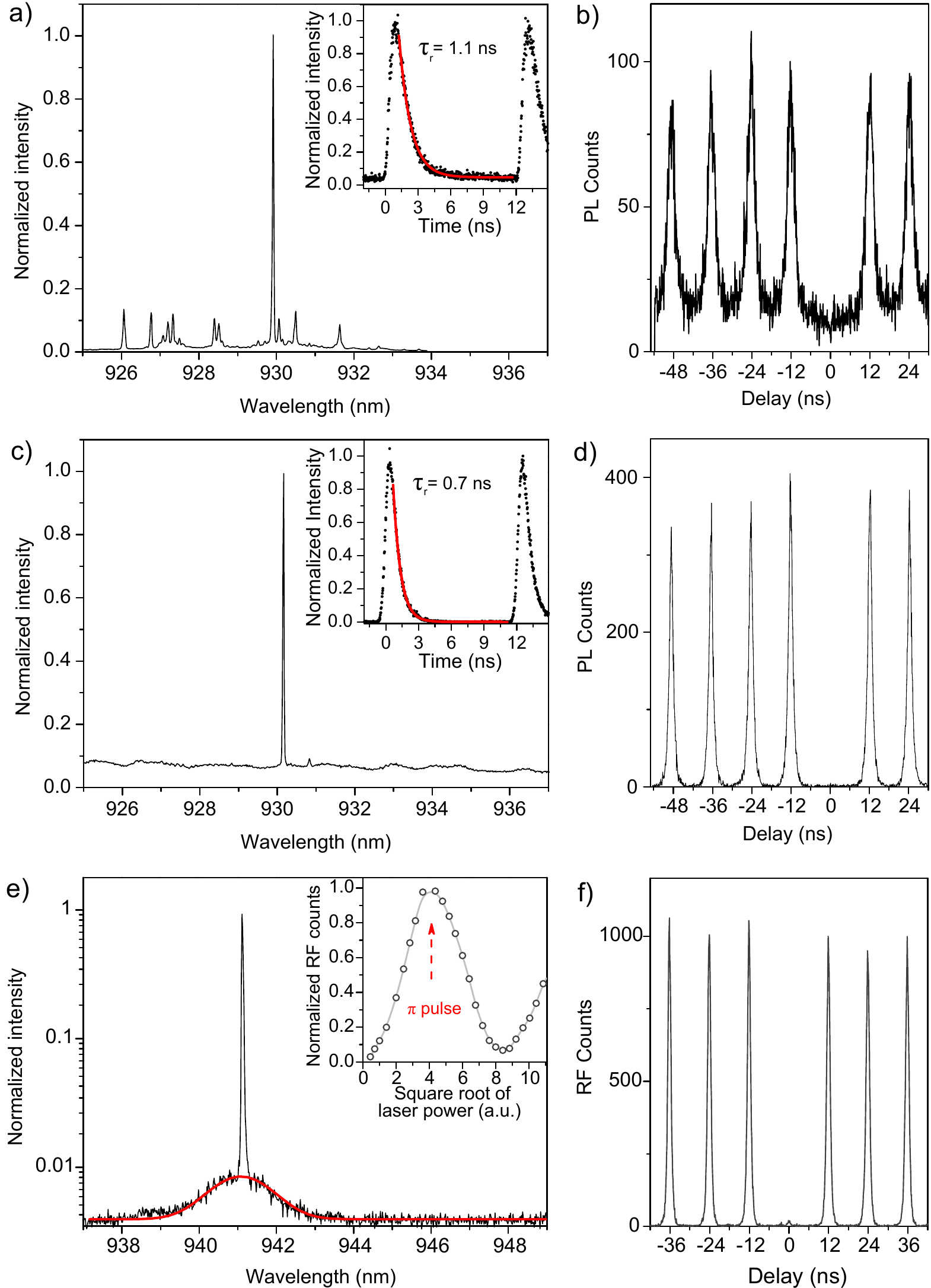}
\caption{\label{Abb:SPS2} a) Photoluminescence (Inset: Time resolved PL) signal of a QD under non-resonant excitation; c) quasi-resonant excitation; e) resonance fluorescence. The inset in e) depicts the power dependency of the emission under pulsed resonance fluorescence, which is characterized by the occurrence of distinct Rabi-oscillations. Corresponding second order correlation histograms: b) non-resonant pumping, d) p-shell excitation; f) resonance fluorescence.).
}
\end{figure}

Fig.~\ref{Abb:SPS2}a) depicts a photoluminescence spectrum recorded under non-resonant excitation condition. The spectrum is characterized by a pronounced emission feature which is attributed to the neutral exciton emission from a single QD via its spectral properties (such as polarization and power dependency). In the close spectral vicinity we detect a number of other lines, possibly stemming from different charge configuration of the same QD or from neighboring QDs. In the following, we will focus on the characteristics from the brightest transition: The corresponding second order autocorrelation function recorded from this line is shown in Fig.~\ref{Abb:SPS2}b), recorded under pulsed excitation with a repetition frequency of 82 MHz. During each excitation pulse, carriers are excited non-resonantly in the wetting layer of the QDs, which is reflected in the distance between the peaks in the correlation histogram. As expected from a single quantum emitter, at $\tau=0$ the coalescence probability is strongly suppressed and a dip occurs in the histogram. The corresponding $g^{(2)}(\tau=0)$ value of 0.05 is a clear signature of single photon emission. 

However, as a result of the non-resonant excitation, a large number of carriers is generated in the surrounding of the QD during one excitation pulse. If the lifetime of this reservoir exceeds the average recombination time of the QD trion, carriers can re-excite the QD after the first recombination event, which leads to a strong broadening of the peaks in the correlation histogram. \\

This effect is largely suppressed when the QD is quasi-resonantly excited into an excited state: In Fig \ref{Abb:SPS2}c) we plot the photoluminescence spectrum of a QD which is excited with a laser detuning of 29 meV on the high energy side of the recombination line. Stray-light from the excitation laser is spectrally suppressed by a combination of bandpass filters. Due to the quasi-resonant nature of the excitation, the spectrum is almost background free and a single bright emission line dominates the spectrum over a wide range. More importantly, the effects of strong time jitter and carrier recapturing are suppressed in the corresponding correlation histogram (Fig \ref{Abb:SPS2}d) and the width of the coincident peaks in the histogram are now determined by the lifetime of the excitonic transition with a characteristic time of 700 ps, which is in good agreement with the time resolved PL trace (inset of Fig \ref{Abb:SPS2}c)). This effective reduction of the measured lifetime (compared to inset of Fig \ref{Abb:SPS2}a) is again consequence of the absence of carrier recapturing under quasi-resonant excitation. 
The purity of the source is even improved, as characterized by a value of $g^2(\tau=0)=0.023$ which was directly extracted from the raw data without any background correction. \\

A truly coherent, time-jitter free excitation method made use of an ultrafast (3 ps) pulsed laser with its central frequency resonant with the QD transition. 
In Fig.~\ref{Abb:SPS2}e) we plot a spectrum of the QD-emission signal under such resonance flourescence (RF) conditions. The narrow emission line stemming from the driven QD resonance sits on top of a broad, yet dim background from the pump laser (plotted in log-scale). The much broader laser background can be further filtered using a narrow-band etalon, resulting in a signal to background ratio exceeding $300:1$ \cite{He-NatNano13}. The QD emission intensity is plotted as a function of the square root of the pump power of the excitation laser in the inset of Fig.~\ref{Abb:SPS2}e): The observed Rabi-oscillations are characteristic for a resonantly driven two-level system and reflect the coherent nature of the excitation process. Single RF photons are deterministically generated at the peak maximum, which corresponds to a $\pi$-pulse in the picture of driven Rabi-oscillations. 
The according second order correlation measurement under these conditions is shown in Fig.~\ref{Abb:SPS2}f). In order to further filter out the broad background from the excitation laser, the emission is fed through a high finesse Fabry-Perot ethalon. Under such configurations, the second order autocorrelation function can reach values as low as $g^2(\tau=0)=0.003$, pointing out the character of our single QD as a almost perfect single photon source \cite{Wei-arxiv14}.

\subsection{Two photon interference with single photons}

The combination of high brightness and high purity of the single photon emission allows us to carry out photon interference measurements with the experimental configuration briefly described in section \ref{section:theory} and  chapter (by Lanco i Senellart). 

First, we study the two-photon interference (TPI) of consecutive photons emitted from a single QD, to scrutinize the dependency of the interference visibility on the excitation condition. Therefore, we adjust the path length difference of the two arms in our Mach-Zehnder interferometer to the laser repetition period of 12.2 ns (see Fig. \ref{Abb:HOM}d), so that two consecutively excited photons can coincide at the same time on the beam-splitter. Additionally, there is the possibility to change the time delay between the two arms of the interferometer via a variable optical fiber delay.

\begin{figure}[t]
\centering
\includegraphics[width=12cm]{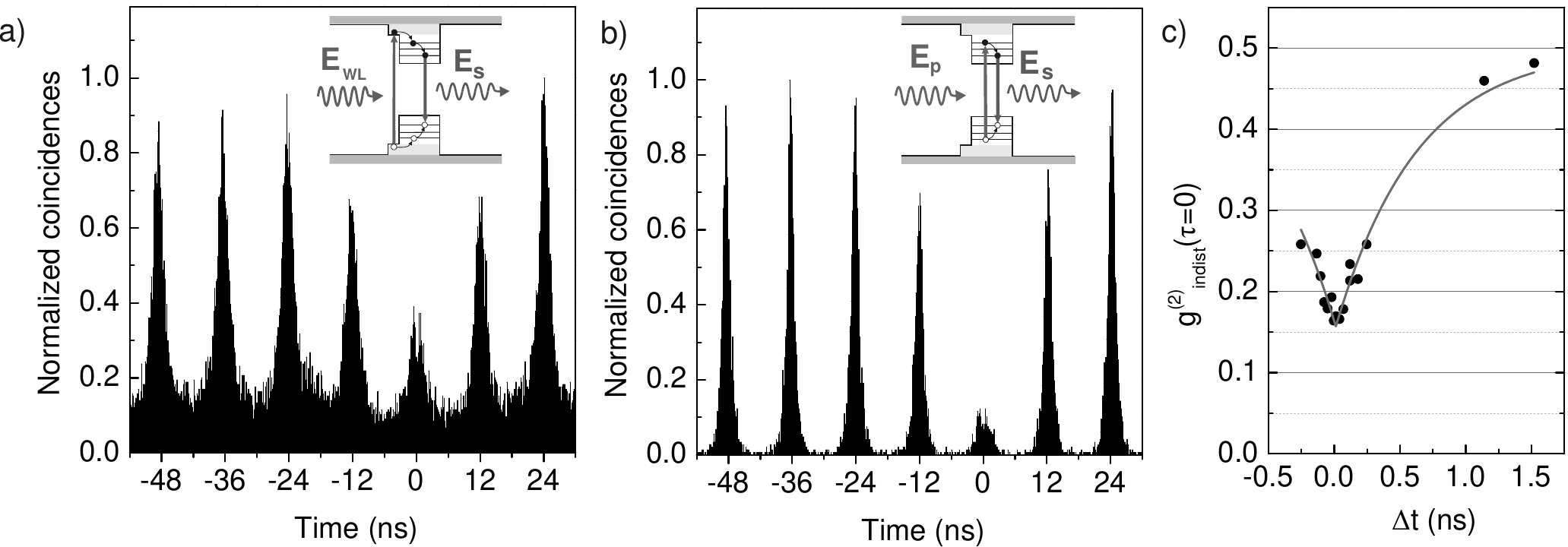}
\caption{\label{Abb:ID1} a) Two photon interference histogram for a QD. The emitter is non-resonantly excited into the wetting layer. b) TPI under p-shell excitation. c) The Hong-Ou-Mandel dip evolves when the TPI visibility is plotted against the delay time $\Delta t$. 
}
\end{figure}

We first study the TPI of consecutive emitted photons under wetting layer excitation. As we have discussed above, the very long diffusion lengths of our sample lead to a recapturing of charge carriers after a first recombination. This recapturing results in a background in the autocorrelation histogram at $|\tau|>0$. The corresponding correlation histogram for the TPI of consecutive emitted photons at an interferometer path length difference of $\Delta t \approx 0$ is shown in Fig. \ref{Abb:ID1}a). Fitting the data with a model based on Ref. \cite{Bylander-EPJD03} yields a visibility of only 12$\%$. This small value is a result of the very large time uncertainty induced by the long emission time induced from the carrier recapturing, which makes a simultaneous collision of two photons on the beam splitter very unlikely. 

As a direct comparison, in Fig. \ref{Abb:ID1}b)  the second order correlation function for TPI is shown for zero path length difference for a QD under p-shell excitation. Here, the peak at $\tau=0$ is strongly suppressed below a value of 0.5. The probability for two photons that coincide at the beamsplitter and exit in opposite directions $g_{indist}^{(2)} (\tau=0)$ is given by the area under the peak at $\tau=0$ divided by the averaged area of four peaks for $|\tau|>\pm12.2$ ns. From the raw data we extract values of $g_{indist}^{(2)}=0.16 <0.5$, verifying the indistinguishability of the photons generated under quasi resonant pumping. 
In order to accurately extract the visibility of the two photon interference, we investigate  $g_{indist}^{(2)} (\tau=0)$  in dependence of interferometer path length offset $\Delta t$. In this manner, we observe the characteristic Hong-Ou-Mandel dip for $\Delta t =0$. 
Via fitting the data with a two sided exponential $g_{indist}^{(2)} (\Delta t)=0.5[1- v e^{(-|\Delta t|/\tau_m )} ]$ we can extract a non-postselected value of TPI visibility $v$ of 69$\%$. This value is comparable to the values of QDs embedded in micropillar cavities \cite{Santori-Nature02, Gazzano-NatCom13}, where the Purcell effect is employed to reduce the radiative decay time $\tau_r$ and hence improve $\frac{\tau_c}{2\tau_r}$. 
The visibility under quasi resonant excitation is strongly increased compared to the non-resonant excitation scheme resulting from the direct excitation conditions which lead to a reduced uncertainty in the emission time, a lack of carrier re-capturing processes and reduced charge carriers in the wetting layer. For an ideal spontaneous-emission source, with instantaneous initial excitation and no decoherence, $\tau_m$ which characterizes the arrival time of photons on the beam splitter would be equal to the spontaneous emission lifetime of the quantum emitter. From the fit we get values of $\tau_m = 630$ ps which indeed is close to the spontaneous emission lifetime of this QD (see inset Fig. \ref{Abb:SPS2}c)).

\begin{figure}[h]
\centering
\includegraphics[width=12cm]{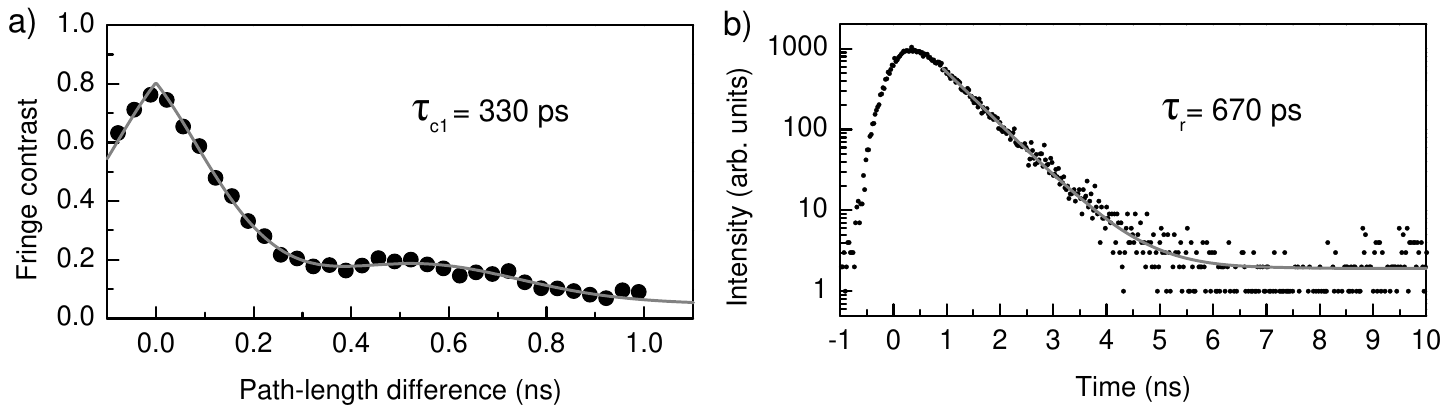}
\caption{\label{Abb:T2} a) Coherence time and b) decay time of the QD characterized in Fig~\ref{Abb:ID1}.
}
\end{figure}

Ideally, in order to obtain maximum degrees of indistinguishability, Fourier-transform limited sources are required. As we have assessed in \ref{section:theory}, the relation between coherence time $\tau_c$ and lifetime of the QD emission $\tau_r$, that describes the visibility of TPI is $\nu_{max}=\frac{\tau_c}{2\tau_{r}}=1$. In order to directly assess the coherence time of the QD emission under p-shell excitation, we use a Michelson interferometer, and measure the photon interference signal as a function of the variable time delay (see \cite{Michler-2} for details on the method). The fringe contrast in dependence of interferometer path length difference is shown in Fig. \ref{Abb:T2} a). The fine structure splitting of the neutral exciton line leads to oscillations in the interference fringe contrast. From a fit (red solid line) to the experimental data with the Fourier transform of two Lorentzians we can extract the coherence times of the fine structure split lines $\tau_{c1}=330$ ps and $\tau_{c2}=180$ ps.  Compared with the extracted decay time $\tau_r=670$ ps we can extract a maximum visibility for TPI in this case of 25$\%$.

This seeming discrepancy to the extracted value of 69 $\%$ has been observed before \cite{Santori-Nature02} and can be explained by a primary inhomogeneous broadening of the emission lines, for example by charge fluctuations in the vicinity of the QD. These fluctuations can take place on a timescale much longer than the laser repetition frequency, and are hence not affecting the TPI measurement, since only the interference from consecutive photons is measured. On the other hand, in the time averaged Michelson experiment, the interference of photons emitted at much larger time delays contribute, which explains the reduced coherence times. As we will show in the following chapter, however, this argument is no longer true if photons from independent sources are interfered, which asks for the capability to generate photons close to the Fourier limit.

\begin{figure}[h]
\centering
\includegraphics[width=12cm]{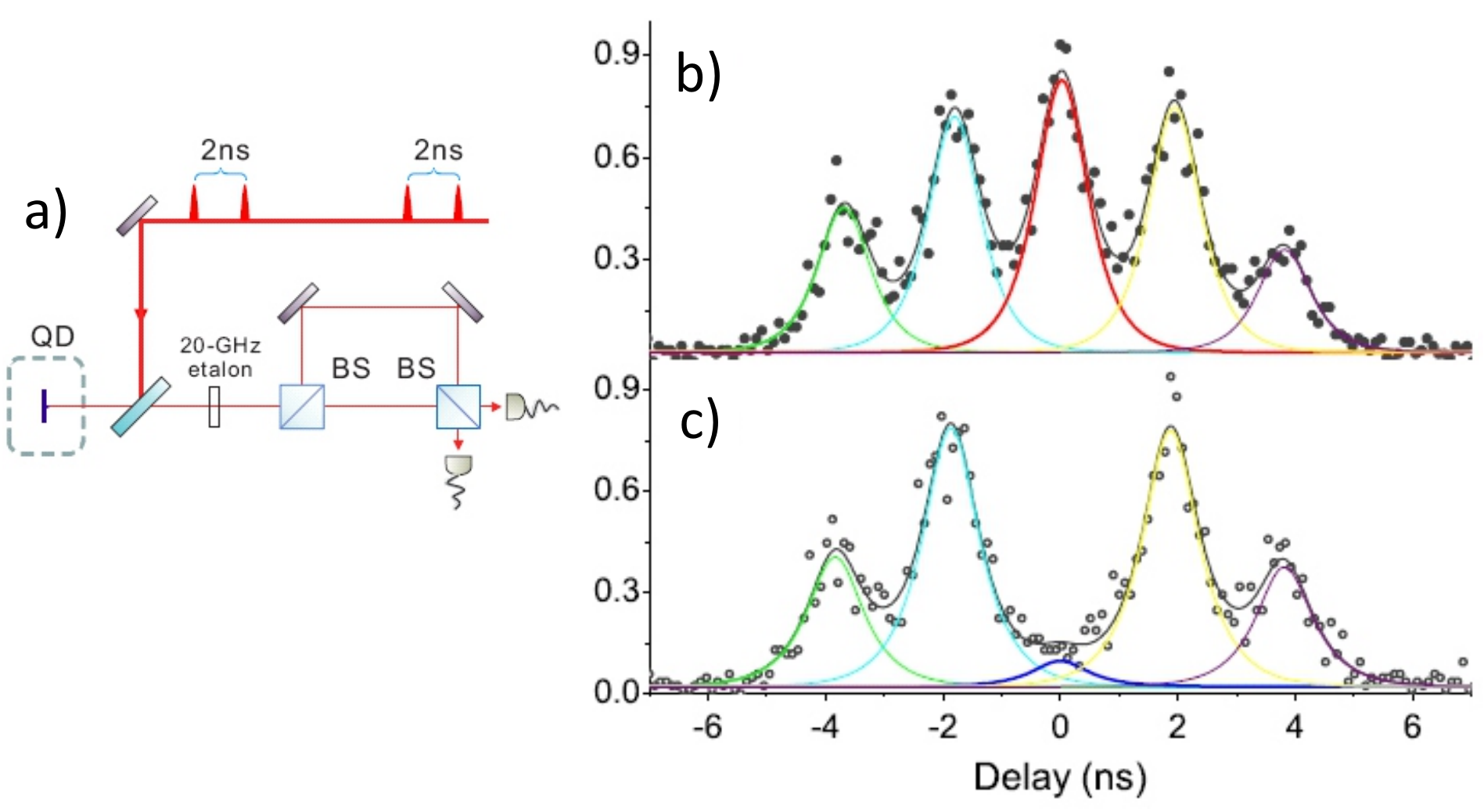}
\caption{\label{Abb:RF} Photon interference spectroscopy in the resonance fluorescence configuration: a) Sketch of the optical setup: Each pulse of the excitation laser is split into two pulses with a time delay of 2 ns, resulting in a two-fold excitation of the same QD. The fluorescence is fed into a Hong-Ou-Mandel setup, and the RF-photons are recombined on the second beamsplitter. b) Interference histogram with photons of perpendicular polarizations and c) photons of the same polarization. The absence of the peak at $\tau=0$ demonstrates highly indistinguishable photons.
}
\end{figure}

To generate such photons and increase the visibility of the TPI, we study a QD under pulsed resonance fluorescence conditions. In this experiment, each excitation pulse of the pumplaser is split into two pulses with a delay of 2 ns, generating two single photons each 12.5 ns (see Fig~\ref{Abb:RF} a). The according correlation histogram from the TPI is depicted in Fig.~\ref{Abb:RF}b) and c). If we combine photons with opposite polarizations on the last beam splitter, we observe the correlation histogram of two perfectly distinguishable photons depicted in Fig~\ref{Abb:RF}b), featuring a central peak at $\tau=0$ with the same magnitude as the neighboring peaks stemming from photons with a time difference of 2 ns. If photons with the same polarization are combined on the beam splitter (Fig.~\ref{Abb:RF}c)) the strong suppression of the central peak indicates the high degree of indistinguishability of the photons generated under these conditions. By evaluating the areas under the coincidence peaks, we can directly extract a raw TPI visibility as high as 91 $\%$, clearly exceeding the value for quasi-resonant excitation. A more recent experiment used adiabatic rapid passage to deterministically and more robustly generate single photons and demonstrated a new record of two-photon interference raw visibility of about 98$\%$ \cite{Wei-arxiv14}.
The strong increase of the TPI for resonance fluorescence clearly underlines the superiority of this excitation scheme, and points towards the possibility to deterministically generate single photons close to the Fourier-limit. 

\section{Two photon interference from remote, single quantum dots}
\label{section:experimentII}

\begin{figure}[h]
\centering
\includegraphics[width=12cm]{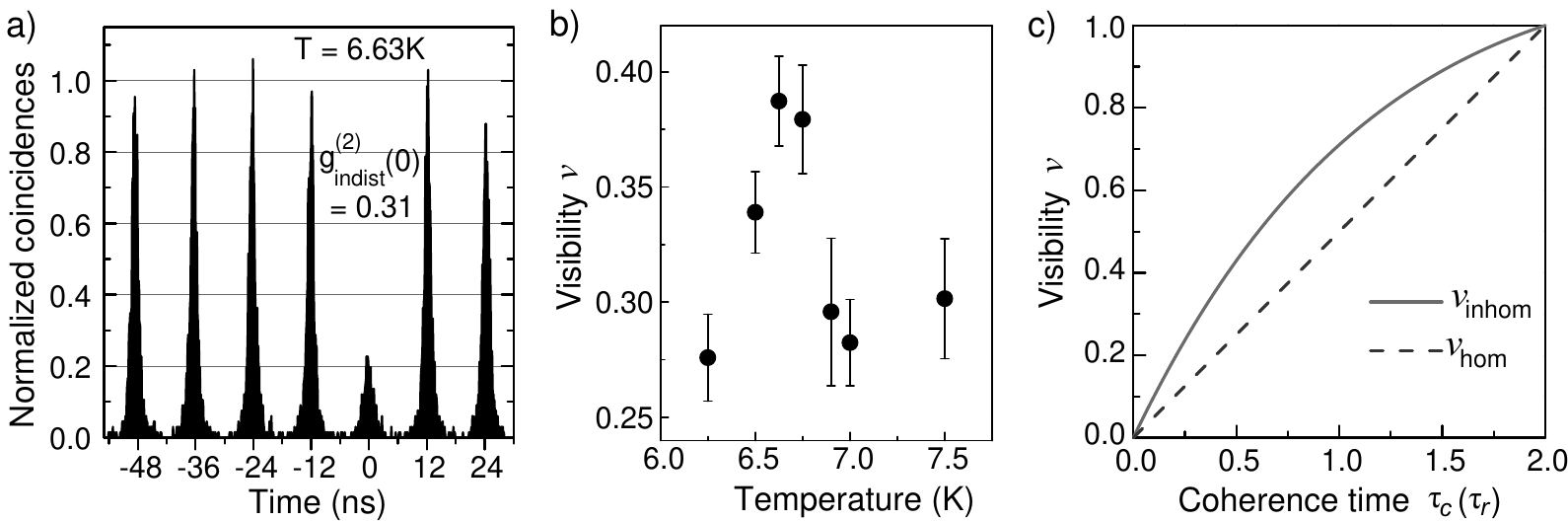}
\caption{\label{Abb:TPI} Two photon interference of QDs emitted from separate sources. The QD samples are mounted in separate cryostats with an overall spatial separation of 0.5 m. a) Interference histogram under spectral resonance. The QDs are tuned into resonance via adjusting the sample temperature. b) Dependency of the two photon interference visibility on the spectral detuning of the QDs. c) Theoretical maximum of the indistinguishability as a function of the QD coherence time (in relation to the decay time $\tau_r$), under consideration of homogeneous (dashed line) and inhomogeneous (solid line) broadening mechanisms.
}
\end{figure}

The very high TPI visibilities which we have discussed in the previous section puts the observation of pronounced TPI effects from photons emitted from separate sources clearly within reach. Such interference effects have previously been observed under non-resonant exictation conditions \cite{Flagg-PRL10} \cite{Patel-NatPhot10}, or under CW resonant fluorescence excitation resulting in time post selecting \cite{Konth-PRL12}. For realistic applications, the non-postselected value of the TPI is however of greater importance, which can only be probed under pulsed excitation conditions. In order to carry out this experiment, we have installed two QD samples in separate cryostats and identified individual QDs with similar emission frequencies and coherence properties as the QD characterized in Fig.~\ref{Abb:ID1}c). Both QDs were excited quasi-resonantly with the same pulsed excitation laser, and the emission was combined on a polarizing beam splitter before it was fed into the Mach-Zehnder interferometer (see. Fig. \ref{Abb:HOM}a). 
In order to probe the two photon interference from separate QDs, the emitters have to be tuned to spectral resonance. A number of tuning mechanisms are possible, including electric fields \cite{Patel-NatPhot10}, strain \cite{Flagg-PRL10} or magnetic fields \cite{Reitzenstein-PRL09}. In our experiments, we utilize the sample temperature to spectrally tune the QD energies to resonance. Since the temperature in the cryostats cannot be varied in a wide range without detrimentally affecting the QD emission properties, a pair of QDs was selected with an energy difference as small as 3 $\mu$eV. The emission energies could then be equalized by changing the temperature of one sample by only 1.8 K, which should only minorly affect coherence properties of the emission.

The second order correlation function for TPI from separate sources is shown in Fig. \ref{Abb:TPI}a) for QD1 at 5.00 K and QD2 at 6.75 K. We determine the opposite output probability $g_{indist}^{(2)} (\tau=0)$ from the raw data by the area of the peak at $\tau=0$ divided by the averaged area of 6 peaks for $|\tau|>0$. From the data we extract $g_{indist}^{(2)} (\tau=0)=(0.31 \pm 0.01)$, which verifies that the photons from the two QDs have a nonzero coalescence probability. The according TPI visibility amount to $v=(39\pm2)\%$, which is the highest value of non-postselected two photon interference from separate QDs observed so far under quasi- or nonresonant excitation conditions. By varying the sample temperature in one cryostat, we can tune the QDs out of resonance, which is directly reflected in a reduced two photon interference visibility, as depiced in Fig.\ref{Abb:TPI}b). 

We will now compare these experimentally observed values with a theory which only takes into account pure dephasing as a decoherence mechanism. As we have described in section \ref{section:theory}, in the presence of pure dephasing limiting the coherence time $\frac{1}{\tau_c}=\frac{1}{2\tau_r}-\frac{1}{\tau_{deph}}$ the maximum visibility of the TPI is obtained for $v = \frac{\tau_c}{2\tau_r}$. Taking into account the experimentally extracted radiative decay time of 670 ps (Fig. \ref{Abb:T2}b) and coherence time of 330 ps (Fig. \ref{Abb:T2}a), we can infer a maximum interference visibility of 25 $\%$, which is clearly exceeded in our experiment. For two photons emitted from the same QD, we argued that frequency jitter on a time scale beyond the repetition time of the pulsed excitation laser led to a reduction of the coherence time which was however only weakly affecting the interference visibility from consecutive photons. For photons emitted from independent sources, clearly this argumentation is not valid. In order to take account for the effects of inhomogeneous broadening in the evaluation of the TPI, various frequency components have to be taken into account. We follow the analysis by Legero et al. (see chapter (Kuhn, Zhao) of this book and \cite{Legero-AAMOP06}) to derive an expression for the visibility of TPI in the presence of inhomogeneous broadening represented by a Gaussian frequency distribution. We assume that the two photons which interfere at the beam splitter originate from independent ensembles of Fourier transform limited photons. The Fourier limited single-photon wave packets for QD1 and QD2 are one sided exponential functions: 
\begin{equation}
\xi_1(t)=\begin{cases} \sqrt[4]{\frac{1}{\pi\tau_r}}e^{-\frac{t-\frac{\delta\tau}{2}}{2\tau_r}-i\left(\omega-\frac{\Delta}{2}\right)t}  & \text{if }t-\frac{\delta\tau}{2}>0\\
  0 &\text{otherwise}
\end{cases}
\end{equation}
and
\begin{equation}
\xi_2(t)=\begin{cases} \sqrt[4]{\frac{1}{\pi\tau_r}}e^{-\frac{t+\frac{\delta\tau}{2}}{2\tau_r}-i\left(\omega+\frac{\Delta}{2}\right)t}  & \text{if }t+\frac{\delta\tau}{2}>0\\
  0 &\text{otherwise}
\end{cases},
\end{equation}
where $\delta\tau$ is the time delay and $\Delta$ the frequency difference between them. An inhomogeneous broadening of the emission lines can be considered by a Gaussian frequency distribution $f(\omega)$ with $\sigma^2$ being the variance:
\begin{equation}
f_{i=1,2}(\omega_i)=\frac{1}{\sqrt{\pi} \, 2 \sigma_i}e^{-\frac{\left(\omega_{0i}-\omega_i\right)^2}{2 \sigma_i^2}}.
\end{equation}
With $\omega_1=\omega$ and $\omega_2=\omega+\Delta$ we get the frequency distribution as a function of the frequency difference $\Delta$ 
\begin{equation}
f(\Delta)=\int\mathrm{d}\omega \, f_1(\omega)f_2(\omega,\,\Delta)=\frac{1}{\sqrt{\pi} \, 2\sigma_g}e^{-\frac{\left(\Delta-\Delta_0\right)^2}{4\sigma^2_g}},
\end{equation}
with $\sigma_g=\sqrt{\sigma_1^2+\sigma_2^2}$ and $\Delta_0=\omega_{02}-\omega_{01}$. The correlation function is then given by
\begin{equation}
G_{\mathrm{inhom}}^{(2)}(t_0,\,t_0+\tau)=\int\mathrm{d}\Delta \, f(\Delta)\,\mathrm{tr}\!\left(\hat{\rho}(\xi_1,\xi_2) \, \hat{A}(t_0,\,t_0+\tau)\right),
\end{equation}
where $\mathrm{tr}\!\left(\hat{\rho}(\xi_1,\xi_2) \, \hat{A}(t_0,\,t_0+\tau)\right)$ is the correlation function for two Fourier transform limited  photons with the same polarization: 

\begin{equation}
G_{\mathrm{TL}}^{2}(t_0,t_0+\tau)=\frac{|\xi_1(t_0)\xi_2(t_0+\tau)-\xi_2(t_0)\xi_1(t_0+\tau)|^2}{4}
\end{equation}

Finally, the probability for detecting a photon at time $t_0 + \tau$ in one output of the beam splitter while a photon is detected at time $t_0$ in the other one for inhomogeneous
broadened ensembles of photons is given by
\begin{align}
P_{\mathrm{inhom}}&=\int^{\infty}_{-\infty}\!\mathrm{d}t_0 \, G_{\mathrm{inhom}}^{(2)}(t_0,\,t_0+\tau)= \nonumber\\
&=\frac{1}{8\tau_r}\left(e^{-\frac{\left|\delta\tau-\tau\right|}{\tau_r}}+e^{-\frac{\left|\delta\tau+\tau\right|}{\tau_r}}-2\cos\!\left(\Delta_0\,\tau\right)e^{-\frac{\left|\delta\tau\right|+\left|\tau\right|}{\tau_r}}\,e^{-\sigma_g^2\tau^2}\right)
\end{align}

From this expression, we yield an expression for the two photon interference visibility for $\delta \tau=0$ and $\Delta_0=0$, only depending on the radiative decay time $\tau_r$ and the geometric average of the broadening of the two photon ensembles $\sigma_g=\sqrt{\sigma_1^2+\sigma_2^2}$:
\begin{equation}
v_{\mathrm{inhom}}=1-\frac{1}{\tau_r \sigma_g} \left( 2 \tau_r \sigma_g - e^{\frac{1}{4\tau_r^2 \sigma_g^2}} \sqrt{\pi} \mathrm{erfc}\left(\frac{1}{2 \tau_r \sigma_g}\right)\right)
\label{eq:vis}
\end{equation}

In Fig. \ref{Abb:TPI}c) we plot the maximum TPI of such inhomogeneously broadened wave packets as a function of the coherence time (in multiples of the lifetime $\tau_r$), to visualize the strong influence broadening's origin on the maximum interference visibilty. From this analysis, we can estimate a maximum visibility of $\nu_{inhom}=(36.4\pm 1.5)\%$ which is in good agreement with the experiment. 
This underlines the importance of effects as spectral wandering, time jitter and other inhomogeneous broadening channels in particular for photon interference experiments from independent sources. It is worth noting, that major improvements have recently been accomplished by utilizing resonance fluorescence conditions in such an experiment. Due to the suppression of inhomogeneous broadening effects under strict resonant excitation, two photon interference visibilities beyond $80\%$ \cite{Gao-NatCom13, He-PRL13} could be obtained. 

\subsection{Conclusion}

Single semiconductor quantum dots have been established as compact single photon sources on a solid state platform. Towards the implementation of these quantum emitters as sources of highly indistinguishable photons, which is key to realize quantum teleportation schemes and highly desired quantum repeaters, the degree of indistinguishability of the photon emission is a key parameter. As we have reviewed in this chapter, besides utilizing the effects of cavity quantum electrodynamics to modify the radiative decay time of the photon emission, the appropriate excitation scheme plays a crucial role to realize high degrees of indistinguishabilities. In particular resonance fluorescence conditions can be considered as a reliable technique to generate single photons near unity indistinguishability. We highly anticipate that a combination of such sophisticated pumping schemes and the exploitation of light matter coupling effects can lead to even further simultaneous improvements of the photon coupling efficiencies and degrees of indistinguishability, which makes single QDs an truly appealing alternative to cold atoms and ions towards the realization of quantum repeaters.

\

\section{Acknowledgements} %
\begin{acknowledgement}
The authors acknowledge the great support of the following persons throughout the last years: 
S. Maier, A. Thoma, Y. He, Y.-M. He, N. Gregersen, J. Mork, J. Schary, M. Lermer, M. Wagenbrenner, L. Worschech, S. Reitzenstein and A. Forchel. We acknowledge financial support by the BMBF (Projects QuaHLRep and Q.com-H) as well as the state of Bavaria. 
\end{acknowledgement}

\section{References}

\bibliographystyle{SpringerPhysMWM} 
\bibliography{Literatur2}

\printindex
\end{document}